# Cryogenic photonic link using an extended-InGaAs photodiode and short pulse illumination towards high-fidelity drive of superconducting qubits

Takuma Nakamura,[1,2] Dahyeon Lee,[1,2] Jason Horng,[1] Florent Lecocq,[3] John Teufel,[2,3] and Franklyn Quinlan[1,4,*]

[1]*Time and Frequency Division, NIST, Boulder, CO, USA.*
[2]*Department of Physics, University of Colorado Boulder, Boulder, CO, USA*
[3]*Applied Physics Division, National Institute of Standards and Technology, Boulder, CO, USA*
[4]*Electrical, Computer and Energy Engineering, University of Colorado Boulder, Boulder, CO, USA*
*\* franklyn.quinlan@nist.gov*

**Abstract:** We investigate short pulse illumination of a high-speed extended-InGaAs photodiode at cryogenic temperatures towards its use in control and readout of superconducting qubits. First, we demonstrate high detector responsivity at 1550 nm illumination at 20 mK, a wavelength band unavailable to cryogenic standard InGaAs detectors due to the temperature-dependent bandgap shift. Second, we demonstrate an improved signal-to-noise ratio (SNR) at the shot noise limit for cryogenic short optical pulse detection when compared to conventional modulated continuous-wave laser detection. At 40 µA of photocurrent and a detector temperature of 4 K, short pulse detection yields an SNR improvement of 20 dB and 3 dB for phase and amplitude quadratures, respectively. Lastly, we discuss how short pulse detection offers a path for signal multiplexing, with a demonstration of simultaneous production of microwave pulses at two different carrier frequencies. Together, these advancements establish a path towards low noise and power efficient multiplexed photonic links for quantum computing with a large number of superconducting qubits.

## 1. Introduction

Quantum information systems based on superconducting (SC) qubits are a viable path towards large-scale quantum computing, owing to lithographic production of sizeable qubit networks combined with continuous improvements in gate and readout fidelity [1,2]. As these systems continue to increase to greater numbers of physical qubits, there is a growing need to implement large-scale qubit drive and state readout. SC qubits rely on cryogenic delivery of carefully shaped microwave pulses to control and readout the quantum state of each qubit. Presently, these pulses are generated at room temperature and routed through heavily filtered and attenuated coaxial transmission lines to the qubits held at ~10 mK temperatures. In order to reach the million-qubit scale required for truly transformational quantum computing [3], one is confronted with the monumental task of managing the heat load, cost and complexity of millions of coaxial microwave signal lines. Perhaps the most fundamental of these challenges is the heat load, where the cooling power of 20 µW of a typical dilution refrigerator would limit the number of coaxial lines to a few thousand [4]. The clear need for an alternative architecture has led to the exploration of other means of large-scale qubit drive and readout, including optical interconnects, single flux quantum drive and cryogenic CMOS [5–9].

Optical interconnects offer a compelling scaling solution to achieve qubit numbers well beyond what is possible with coaxial cabling. Found wherever large amounts of information need to be transferred at high rates, optical interconnects provide large intrinsic bandwidth, long reach, and low noise. When applied to the control and readout of superconducting qubits, optical interconnects additionally leverage the extremely low thermal conductivity of optical fibers (about 1000x less than coaxial cable), enabling the connection of a million or more qubits without a significant passive heat load. An initial demonstration has shown the viability of this approach, where an optical interconnect delivered control and readout pulses to a single 3D transmon qubit [5].

To advance cryogenic optical interconnects beyond proof-of-principle demonstration and towards large-number scaling, several advancements are desirable. First, to take advantage of telecommunications investments in lasers, modulators, and other optical components, cryogenic photodetectors that operate in a wide wavelength range that extends into the telecommunications C-band (1530 nm to 1565 nm) and L-band (1565 nm to 1625 nm) are needed. Due to the bandgap shift to shorter wavelengths at cryogenic temperatures, this will require a move away from the standard InGaAs detectors that are the workhorse of C- and L-band telecommunications. Second, signal multiplexing schemes will need to be developed that can deliver multiple microwave control and readout signals while minimizing the required optical power and associated heat load. As discussed further below, this implies a departure from direct modulation of a CW laser that was demonstrated in [5]. Third, the signal-to-noise ratio (SNR) of each photonically delivered microwave signal should be optimized. With photodetection comes photocurrent shot noise that reduces the SNR compared to the coherent state of the microwave field. This noise can cause errors in qubit drive and readout in excess over attenuated coaxial lines. As this noise is proportional to the optical power, modulation schemes that deliver the highest microwave power for a given optical power are preferred.

Here we work towards these goals by combining short optical pulse detection for reduced heat load and improved SNR with an extended-InGaAs photodiode at cryogenic temperatures for operation in the optical C- and L-bands. We demonstrate high responsivity at temperatures of 4 K and 20 mK with 1550 nm illumination, representing, to the best of our knowledge, the first experimental demonstration of the utility of extended-InGaAs for 20 mK applications of photonically generated microwave signals. Furthermore, we use short pulse illumination to generate a multiplicity of microwave carriers at 20 mK, any of which can be selected and used for qubit drive or readout. Importantly, generation of several microwave carriers with short pulse detection requires less optical power than generating just a single carrier with modulated-CW laser light. We then explore the advantages of short pulse detection over the detection of modulated-CW laser light in terms of SNR, with careful measurement of the shot noise-limited amplitude and phase noise of a microwave carrier. For the same average photocurrent, we show an improvement of 20 dB in the microwave phase noise when using short pulse detection compared to modulated-CW detection. This is the largest phase noise difference demonstrated to date, either cryogenic or room temperature. Finally, we discuss how multiple microwave tones may be utilized simultaneously for multiple qubits from the detection of a single train of optical pulses, with a demonstration of simultaneous generation of microwave pulses at carrier frequencies of 1 GHz and 3 GHz.

## 2. Experimental results

*Responsivity and Power.* First, we verified the operation of an extended-InGaAs detector at cryogenic temperatures. We used a fiber-coupled extended-InGaAs detector with 2.2 μm bandgap and 10 GHz bandwidth [10,11]. We measured the responsivity at both room and cryogenic temperatures using a purpose-built, broadly tunable light source, in which one color of supercontinuum light was extracted by a grating. For comparison, a standard InGaAs photodiode was also tested at both room temperature and 4 K. As shown in Fig. 1(a), both photodiodes show expected bandgap shifts [12] with decreasing responsivity at longer wavelengths. As a result, at cryogenic temperatures, the standard InGaAs detector is usable only below ~1500 nm. On the other hand, the extended-InGaAs detector still has a high responsivity of 0.7 A/W from 1300-1900 nm, covering most optical telecommunications bands.

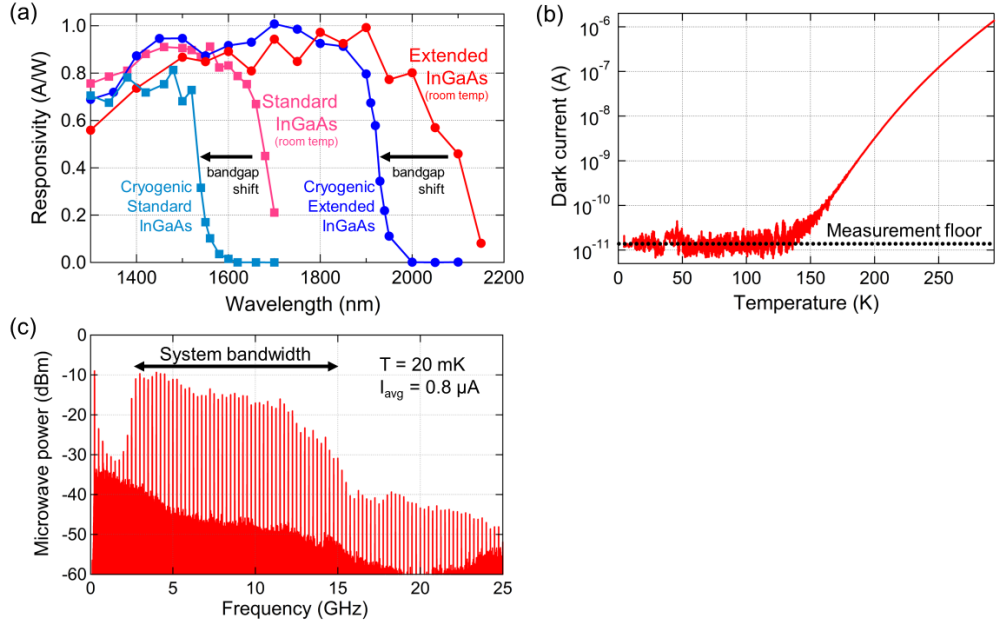

Fig.1 (a) Responsivities for the extended and the standard InGaAs photodiodes at room temperature and cryogenic temeprature. The bandgap shifts are observed. (b) Temperature dependence of the dark current of the extended InGaAs photodiode. Applied reverse bias is 5 V. (c) Measured microwave spectrum from extended-InGaAs photodiode at 20 mK illuminated by 250 MHz mode locked laser, generating 0.8 µA average photocurrent. The reported power level is after several stages of amplification. Filters and isolators in the coaxial cabling from the cryogenic photodiode to the room-temperature spectrum analyzer constrained the system measurement bandwidth.

A typical drawback of extended-InGaAs is higher dark current than that of standard InGaAs, caused by defects from the lattice constant mismatch with the InP substrate [13]. However, photodetector dark current decreases exponentially as the temperature decreases, and is expected to be negligibly low at cryogenic temperatures. To verify this with the extended-InGaAs photodetector, we measured the temperature dependence of its dark current at –5 V bias. As shown in Fig. 1(b), for temperatures below 120 K, the dark current falls below the 10 pA noise floor of the measurement.

Measurements of responsivity and dark current were followed by measurements of the RF spectral response of the cryogenic extended-InGaAs detector under ultrashort optical pulse illumination. An example spectrum is shown in Fig. 1(c). As an optical pulse source, we used a mode-locked laser whose repetition rate is 250 MHz. Measurements were made with the detector at ~20 mK and an average photocurrent of 800 nA. Within the system measurement bandwidth, over 40 tones are generated on a 250 MHz grid. The power in each microwave tone is given by

$$P_\mu = 2I_{avg}^2 R|H(f)|^2, \qquad [Eq.(1)]$$

where $I_{avg}$ is the average photocurrent, $R$ is the system impedance, and $H(f)$ is the transfer function of the electrical circuit, including the electrical response of the photodiode [14]. We contrast this array of tones with the single tone generated by modulating a CW laser, where the photodetected microwave power is given by

$$P_\mu^{CW} = \frac{1}{2}I_{avg}^2 R|H(f)|^2 \qquad [Eq.(2)]$$

Thus, not only is the output limited to one microwave tone, but the microwave power of this tone is 4x lower for a given average optical power and photodiode response than the power attained in each tone with short pulse detection. More discussion on how we envision exploiting multiple tones for qubit drive and readout from a single optical pulse train is presented in Section 3.

*Amplitude and Phase Noise.* The increased microwave power for a given average photocurrent under short pulse illumination immediately confers an SNR advantage over modulated-CW detection. Importantly, this improvement in the SNR is not equally divided between amplitude and phase quadratures of the generated microwave carrier [15,16]. Under short pulse illumination, photogenerated carriers are localized within a very short time window, leading to a reduced photocurrent shot noise contribution to the microwave timing, or phase, noise. Therefore, the phase noise of the derived microwave signal can be orders of magnitude lower than the shot noise-limited amplitude noise. The separation between amplitude noise and phase noise is optical pulse width dependent, with the shot-noise limited phase noise on a 1 GHz carrier more than 30 dB below the amplitude noise for 1 ps optical pulses. However, room temperature phase noise measurements of optically derived microwaves have not reached this limit, and are believed to be limited by photocarrier scattering off lattice phonons [17]. As the level of scattering is expected to be device and temperature dependent, noise measurements of the extended-InGaAs detector at cryogenic temperatures are warranted.

The experimental setup for our amplitude and phase noise measurements of a photonically generated microwave carrier using the cryogenic extended-InGaAs detector is shown in Fig. 2. We compared short pulse illumination to sinusoidally modulated CW light under various average photocurrents. As a short pulse source, we used an Er:glass mode-locked laser with center wavelength of 1550 nm and 500 MHz pulse repetition rate [18]. Prior to illuminating the photodiode, the repetition rate of the optical pulse train was doubled to 1 GHz by a free-space pulse interleaver [19]. The inclusion of the interleaver allowed us to increase the average photocurrent with minimal photodiode saturation and electrical pulse distortion. The optical pulse width was measured with an intensity autocorrelator, and was estimated to be less than 1 ps at the detector. For the modulated-CW source, a 1550 nm fiber laser was sinusoidally modulated by a microwave synthesizer driving an electro-optic modulator at a frequency of 1 GHz. Standard telecom single-mode fiber delivered the light to the cryogenic photodetector at 4 K. As before, the applied bias voltage was –5 V.

We used cross-spectrum analysis to extract the phase and amplitude noise on the 1 GHz carrier. In addition to rejecting the noise of the rf amplifiers in the measurement chain, cross-spectrum measurements can reject thermal noise at the detector plane to reveal the underlying photocurrent noise [20]. This was particularly important in revealing the phase noise from photodetection. The signal from the detector was split into two identical amplification branches by a power splitter. The first stage of amplification was at 4 K, allowing us to take advantage of the low noise-temperature of a cryogenic amplifier. To reduce the heat load at the 4 K stage, we utilized long, low thermal conductivity microwave cables, resulting in >5 dB loss at 1 GHz from the 4 K stage to the room temperature environment. This forced us to measure shot noise at average photocurrents above 5 µA, even with cross-correlation and the benefits of cryogenic amplification. Subsequent amplification stages at room temperature provided enough microwave power to saturate the mixers used in the phase and amplitude noise bridge [21]. The separate signal branches were mixed with independent reference oscillators at 1 GHz to down-convert the photocurrent noise to baseband for measurement. Either the amplitude or phase quadrature was chosen for measurement with phase shifters in the reference oscillator path, such that the photodetected signals were either in phase or in quadrature with their reference oscillator. The lasers' repetition/modulation frequencies and reference oscillators were phase-locked with low bandwidth (~100 Hz) to maintain the relative average phase for either amplitude or phase noise measurement. Note that the noise measurements were made at offset frequencies (>10 MHz) well outside the lock bandwidth to ensure the presence of the phase lock did not skew the noise measurement results.

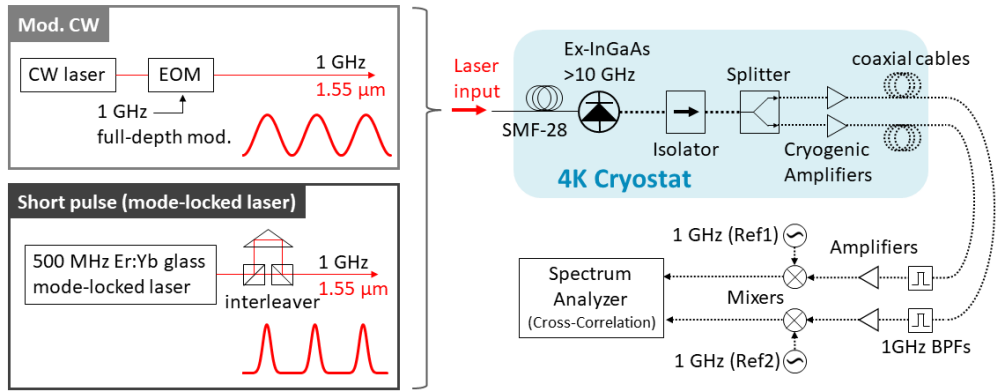

Fig.2 Experimental setup for the cross-spectrum noise floor measurement with different illumination conditions. One of the lasers was chosen to illuminate the photodiode. The repetition or modulation frequencies of lasers and 1 GHz oscillators were loosely phase-locked to maintain the constant phase difference to select either the amplitude or phase quadrature. BPFs: Band Pass Filters

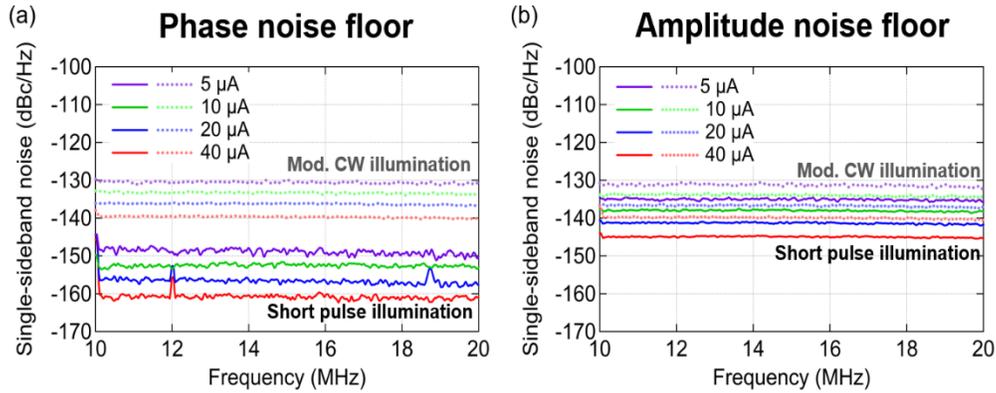

Fig.3 Noise floor measurements with modulated (mod.) CW and short pulse illumination. Phase (a) and amplitude noise floors (b) are roughly equal in the case of modulated-CW. In contrast, short pulse illumination results in a phase noise floor that is much lower than its amplitude noise. For both amplitude and phase quadratures, the noise floor for short pulse illumination is lower than that of modulated-CW illumination.

The results of the phase and amplitude noise measurements are shown in Fig. 3. Measurements are reported for offset frequencies of 10-20 MHz. In addition to residing well outside the lock bandwidth, this offset frequency range is less susceptible to contamination from flicker noise, acoustic noise, and intensity noise of the lasers. Different colors correspond to the different average photocurrents, controlled by varying the average optical power impinging on the photodiode. As shown in Fig. 3(a), for a given average photocurrent the phase noise of the 1 GHz signal generated via modulated-CW light is roughly 20 dB higher than that of short pulse illumination. This separation is a result of the photocurrent shot noise residing primarily in the amplitude quadrature of the photonically generated microwave for short pulses. Amplitude noise measurements confirm this, shown in Fig. 3(b). Note that the amplitude noise of the modulated-CW case is roughly equal to its phase noise, whereas the amplitude noise for short pulse illumination is 15-17 dB higher than its phase noise. Importantly, however, the amplitude noise (defined as relative to the carrier power) from short pulse detection is still ~3 dB lower than that arising from modulated-CW illumination, due to the higher microwave power provided by short pulse detection.

Noise measurement results are summarized in Fig. 4. Both amplitude and phase noise measurements for the modulated-CW illumination are near the predicted shot noise limited value, and scale with average photocurrent as expected. The fact that the measurements are consistently 1-2 dB higher than expected may be explained by slightly less than full depth of modulation on the optical carrier, resulting in slightly lower microwave power for a given average photocurrent. For short pulse illumination, the measured amplitude noise is consistent with nearly full projection of the shot noise onto the amplitude quadrature. The phase noise under short pulse illumination is well below all other noise measurements at all photocurrents tested, with the largest separation at 400 µA. Note that the measured phase noise is well below the thermal noise limit at 4 K, due to thermal noise rejection in the cross-spectrum measurement setup. Temperature gradients, particularly between the photodiode and the first power splitter, will limit the amount of thermal noise rejection [22]. At an average photocurrent of 40 µA, the roughly 17 dB of thermal noise rejection corresponds to a 0.1 K temperature gradient. Such a temperature gradient is reasonable considering the proximity of the power splitter to the amplifiers in our cryostat, and we believe the finite thermal noise rejection is limiting the phase noise at low photocurrents. For comparison, we also performed a measurement with the extended-InGaAs detector at room temperature using short pulse illumination. At room temperature, despite the use of a cross-spectrum measurement, a higher average photocurrent is necessary to overcome the 20 dB higher thermal noise. At room temperature and an average photocurrent of 400 µA (remaining below the saturation limit of the photodiode), the phase noise remains > 10 dB below the shot noise-limited amplitude noise, slightly smaller than the separation seen at cryogenic temperatures. While this is consistent with higher photocarrier scattering at room temperature, more measurements are necessary for a definitive conclusion.

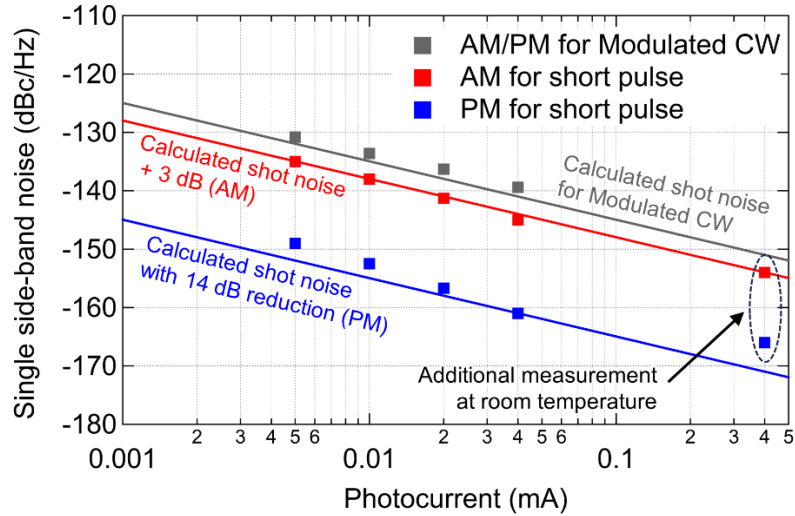

Fig.4 Summary of the noise floor measurements. Additional two points from the same measurement with similar set up at the room temperature were added. Thermal noise at 4 K is calculated assuming short pulse illumination, yielding 6 dB higher SNR than modulated-CW illumination.

## 3. Towards Signal Multiplexing with Short Pulse Detection

In addition to the advantages of higher power and higher SNR, short pulse detection can provide a multiplexing advantage by utilizing the multiplicity of microwave tones generated at the photodetector. Importantly, the use of $N$ microwave tones generated with optical pulse detection requires no more optical power than the generation of a single tone. This is in contrast to signal multiplexing architectures that combines multiple modulated-CW signals on the same photodetector, where the optical power will scale linearly with the number of microwave carriers. In this case, no advantage in terms of heat load is attained, and the increase in the average photocurrent results in higher shot noise for all photonically generated microwaves.

We envision a qubit drive and readout scheme that takes advantage of the multiple tones generated with ultrashort pulse illumination as follows. A train of ultrashort optical pulses, originating from a mode-locked laser, electro-optic frequency comb or microresonator-based optical frequency comb, is passed through an optical pulse shaper. The architecture of the optical pulse shaper can take numerous forms [23–27], but should be able to alter the shape, amplitude, and timing of each pulse. For example, dynamic line-by-line pulse shaping can generate arbitrary optical pulses with an update rate equal to the pulse rate, and pulse-to-pulse timing delays can be controlled [25,26]. A group of pulses is then gated to create a pulse burst, the duration of which determines the duration of the desired microwave pulse. Therefore, the amplitude and phase of each harmonic generated from photodiodes can be controlled in the optical domain.

With arbitrary amplitude and phase control, microwave signals can be tailored to drive individual qubits. Importantly, the shot noise-limited SNR advantage of each tone should be largely preserved, with the exact noise level depending on the particular shape of the microwave waveform [14]. It is important to note that the utilization of multiple microwave tones from short pulse detection necessitates that separate qubits are driven simultaneously, and that the qubit resonant frequencies are aligned on a grid that matches the microwave harmonics of the photonically generated microwave signal. However, superconducting qubits resonant frequencies can indeed be tuned, and for surface codes that utilize multiple qubits with simultaneous drive [28], this trade-off of signal dexterity with power and noise reduction could still be advantageous.

As an initial demonstration along this path, we generated microwave pulses at two different microwave carriers simultaneously from a single optical pulse train and photodetector. The system setup and results are shown in Fig. 5. We used an AOM as a temporal gate to generate a burst of optical pulses that are then directed to a photodetector. The photodetector output was split, and filters at 1 GHz and 3 GHz selected microwave tones at these frequencies. After filtering, the microwave pulses were directed to a sampling scope for analysis. For this demonstration, the photodetector was operated at room temperature, and no optical pulse shaping was implemented. The measured microwave pulses with 1 GHz and 3 GHz carrier are shown in Fig. 5(c), with the individual carriers temporally resolved in Fig. 5(d).

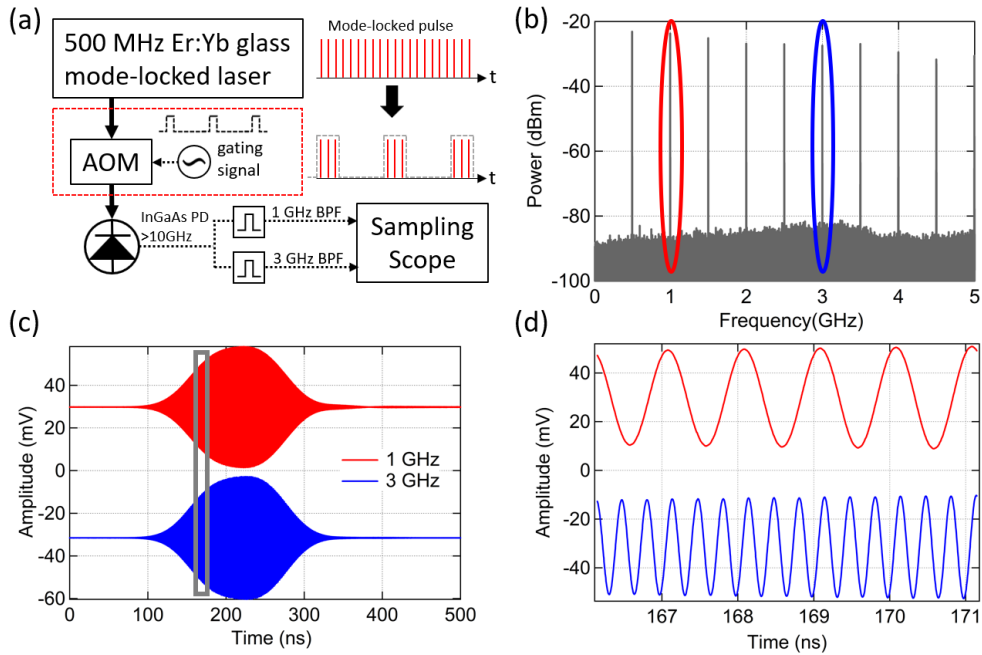

Fig.5 Simple multiplexing experiment. (a) Experimental setup. (b) Microwave tones from detector without gating. Circled tones are selected by filtering. (c) Gated microwave signals measured by the sampling scope with intentional voltage offsets. Enclosed gray area is enlarged in (d).

## 4. Conclusion

We have demonstrated the advantages of extended-InGaAs photodiodes over standard high-speed InGaAs detectors for superconducting circuits and quantum information systems at cryogenic temperatures, including a broader wavelength range that covers important telecommunications bands, high responsivity at 20 mK and sufficiently low dark current. While our focus here has been towards applications at 20 mK, we note that these characteristics of extended-InGaAs would be advantageous for applications at 4 K as well, such as quantum-based voltage standards [27,29], and signal ingress for cryogenic computing [30]. To optimize the SNR of a photonic link for driving or readout of superconducting qubits, we measured the noise floor of photonically generated microwave signals using both sinusoidally modulated CW light and short pulse (~1 ps) illumination. With the same optical power, short pulse illumination gives four times the power in a microwave tone, roughly 20 dB better SNR in the microwave phase quadrature, and 3 dB better SNR in the amplitude quadrature than modulated-CW light. Additionally, we discussed how short pulse illumination results in a multiplicity of microwave tones, and how simultaneous use of these tones can reduce heat load. While more work is required to establish the viability of exploiting the multiple microwave tones generated with short pulse illumination, the higher power and lower noise of even a single microwave tone is advantageous over previous demonstrations utilizing modulated-CW light. Combined with extended-InGaAs photodiodes, we see great potential for photonic drive and readout of quantum information systems with superconducting qubits.

**Acknowledgments:** We thank Charles McLemore and Zachary Parrott for helpful comments. This work is funded by NIST. This work is a contribution of an agency of the U.S. Government and not subject to U.S. copyright.

**Disclosures:** The authors declare no conflicts of interest.

**Data Availability:** Data may be obtained from the authors upon reasonable request.